\documentclass[12pt]{article}
\usepackage{latexsym}
\usepackage{graphicx}
\title{Closed timelike curves, superluminal signals, and ``free will'' in universal quantum mechanics}
\author{Hrvoje Nikoli\'c \\
Theoretical Physics Division, Rudjer Bo\v{s}kovi\'{c} Institute, \\
P.O.B. 180, HR-10002 Zagreb, Croatia \\
{\normalsize e-mail: hrvoje@thphys.irb.hr} \\
\makebox[1in]{} \\
}
\date{\today}
\begin{document}
\maketitle
\begin{abstract}
We explore some implications of the hypothesis that quantum mechanics (QM) is universal, i.e., that
QM does not merely describe information accessible to observers, but that
it also describes the observers themselves. From that point of view, ``free will'' (FW) -- 
the ability of experimentalists to make free choices of initial conditions --
is merely an illusion. As a consequence, by entangling  
a part of brain (responsible for the illusion of FW) with a distant particle,
one may create nonlocal correlations that can be interpreted as superluminal signals.
In addition, if FW is an illusion, then QM on a closed timelike curve can be made consistent
even without the Deutch nonlinear consistency constraint.
\end{abstract}

Keywords: Quantum mechanics; free will; superluminal signal; closed timelike curve

\maketitle


\section{Introduction}

In quantum-information theory, quantum mechanics
(QM) is widely viewed as a purely instrumental tool that describes
information accessible to observers. The observers themselves are not described
by QM, but are viewed as external subjects which may manipulate accessible 
information in various ways. In particular, observers are assumed to have
``free will'' (FW) --  the apparent ability of experimentalists 
to make free choices of initial conditions. 
Some implications of this assumption in QM have been discussed in  
\cite{kochen1,kochen2} (see also \cite{BG,tumulka,hooft} for critiques).

Nevertheless, such an instrumental view is not the only
possible interpretation of QM. In particular, the existence of FW
may not be a fundamental property of observers. Instead, FW
may be an emergent feature resulting from complex dynamics
in the brain. Namely, an observer cannot be aware of all processes in his brain. 
Events determined by causes which he is not aware of may be interpreted 
by his consciousness as being determined by FW, even if the true FW does not exist.
In this way, FW may be merely an illusion.

From the practical point of view, it may seem irrelevant whether FW is a genuine 
ability of experimentalists or merely an illusion. Yet, in this paper we argue that this distinction
may be of a practical relevance. More specifically, 
in Sec.~\ref{SEC2}
we study the possibility that QM is a 
{\em universal} theory, i.e., that QM describes everything, including the human brain.
It turns out that the absence of true FW is a natural consequence of universal QM,
and that the illusion of FW can in principle be explained. Using this insight,
we then discuss two different implications that may be of a practical interest.

First, 
in Sec.~\ref{SEC3} 
we argue that
universal QM contains a theoretical possibility to use nonlocal entanglement
for a sort of superluminal signalization, or more precisely, for the illusion of superluminal 
signalization
that for all practical purposes cannot be distinguished from a true one.
The crucial idea is to use one of the entangled particles to affect in an appropriate way
the part of brain responsible for the creation of the illusion of FW.

Second, 
in Sec.~\ref{SEC4} 
we discuss the implications on QM at closed timelike curves (CTC's).
The most popular approach to QM on CTC is the Deutsch \cite{deutsch} nonlinear
self-consistency constraint. The nonlinearity of the constraint leads to various
unusual effects that cannot be realized within ordinary linear QM, the interest in which 
is increasing \cite{deutsch,razni1,razni2,razni3,razni4,razni5,razni6,razni7,razni8}. 
Recently, some alternatives to the Deutch approach have 
also been proposed \cite{lloyd,wallman}. We show that universal QM also provides a
simple alternative to the Deutch nonlinear consistency constraint. 

Finally, the conclusions are drawn in Sec.~\ref{SEC5}.
 
\section{Universal QM and illusional FW}
\label{SEC2}


The idea of universal QM is that there is no fundamental difference
between ``microscopic'' quantum systems, ``macroscopic'' classical systems, and 
observers. Instead, at the fundamental level, everything obeys the quantum laws of physics
in universal QM. In particular,
a measurement can be described by the von Neumann measurement scheme,
in which measurement is nothing but entanglement between 
the measured system and the measuring apparatus.  
For example, if the measured system is in the spin-$\frac{1}{2}$ state
$|\!\uparrow\rangle$ and if the role of the measuring apparatus 
is played by the brain, then a non-demolition measurement of the state $|\!\uparrow\rangle$
can be described by a unitary evolution
\begin{equation}\label{e1}
 |\!\uparrow\rangle \, |{\rm brain \; ready}\rangle \rightarrow 
|\!\uparrow\rangle \, |{\rm brain \; observes \; up}\rangle ,
\end{equation}
where $ |{\rm brain \; ready}\rangle$ is the initial state of the
brain and $|{\rm brain \; observes \; up}\rangle$ is its final state.
Similarly, if the measured system is in the state 
$|\!\downarrow\rangle$, then the non-demolition measurement
is described by
\begin{equation}\label{e2}
 |\!\downarrow\rangle \, |{\rm brain \; ready}\rangle \rightarrow 
|\!\downarrow\rangle \, |{\rm brain \; observes \; down}\rangle .
\end{equation}
If the initial state is in the superposition $ (|\!\uparrow\rangle + |\!\downarrow\rangle )$,
then (\ref{e1}), (\ref{e2}), and the superposition principle imply
the unitary evolution
\begin{eqnarray}\label{e3}
& 
\;\;\;\;
 (|\!\uparrow\rangle + |\!\downarrow\rangle ) |{\rm brain \; ready}\rangle 
& 
\nonumber \\ 
& 
\rightarrow
|\!\uparrow\rangle \, |{\rm brain \; observes \; up}\rangle 
\nonumber \\
& 
\;\;\;\;\;\;
+
|\!\downarrow\rangle \, |{\rm brain \; observes \; down}\rangle .
&
\end{eqnarray}
Similarly, if the initial state is the
Bell state $( |\!\uparrow\rangle  |\!\downarrow\rangle_{\rm d}+ 
|\!\downarrow\rangle |\!\uparrow\rangle_{\rm d} ) $,
where $|\!\downarrow\rangle_{\rm d}$ and $|\!\uparrow\rangle_{\rm d}$
denote the states of a distant particle entangled with the particle
measured by the brain, then (\ref{e3}) modifies to
\begin{eqnarray}\label{e4}
 & 
\;\;\;\;\;\;\;\;\;\;
( |\!\uparrow\rangle  |\!\downarrow\rangle_{\rm d}+ 
|\!\downarrow\rangle |\!\uparrow\rangle_{\rm d} ) |{\rm brain \; ready}\rangle 
& 
\nonumber \\
&
\rightarrow
|\!\uparrow\rangle |\!\downarrow\rangle_{\rm d} \, |{\rm brain \; observes \; up}\rangle 
&
\nonumber \\
&
\;\;\;\;\;\;
+
|\!\downarrow\rangle |\!\uparrow\rangle_{\rm d} \, |{\rm brain \; observes \; down}\rangle .
& 
\end{eqnarray}
The different brain states $|{\rm brain \; observes \; up}\rangle$ and 
$|{\rm brain \; observes \; down}\rangle$ are macroscopically distinguishable,
so the two terms on the right-hand sides of (\ref{e3}) and (\ref{e4})
can be thought of as two distinguishable branches of the wave function.
The macroscopic distinguishability of the branches significantly helps
to understand why only one of the branches is observed, i.e., how
unitary evolution leads to an apparent ``collapse'' of the wave function.
Such a von Neumann description of quantum measurements (not necessarily with brains) 
plays a crucial role in the theory of decoherence \cite{decoh1,decoh2},
as well as in the many-world \cite{mw1,mw2} and Bohmian \cite{bohm2,holland}
interpretations of QM.

The above description of measurement in universal QM is indeed a well-known result.
But can the appearance of FW be also described by universal QM? Here we propose
a quantum description of the emergence of the illusion of FW, by a mechanism 
 very similar to the von Neumann description of quantum measurements. We assume that the 
``decisions'' of the brain are determined
by external influences from the environment, 
where ``external'' refers to influences which are neither
controlled nor consciously observed by the brain. 
In the first step an appropriate external influence
causes the brain to make a corresponding decision. After that, in the second step 
the brain commands the body to perform the decided action.

For example, if the external influence is described by a quantum state
$ |{\rm up}\rangle$ and if the final action corresponds to 
a preparation of another
quantum system in the state $ |\psi_{\uparrow}\rangle$, then the whole
process can be schematically described by a two-step unitary process
\begin{eqnarray}\label{e5}
&
\;\;\;\;
 |{\rm up}\rangle \, |{\rm brain \; undecided}\rangle \, |\psi_0\rangle 
& 
\nonumber \\
&
\rightarrow 
|{\rm up}\rangle \, |{\rm brain \; decides \; up}\rangle \, |\psi_0\rangle
&
\nonumber \\
& 
\rightarrow
|{\rm up}\rangle \, |{\rm brain \; decides \; up}\rangle \, |\psi_{\uparrow}\rangle .
&
\end{eqnarray}
Here it is assumed that the brain is conscious about the brain states
$ |{\rm brain \; undecided}\rangle$ and $|{\rm brain \; decides \; up}\rangle$,
but that it is not conscious about the external influence $|{\rm up}\rangle$.
Similarly, if the external influence is in the state $|{\rm down}\rangle$, then
we have a similar unitary process
\begin{eqnarray}\label{e6}
&
 |{\rm down}\rangle \, |{\rm brain \; undecided}\rangle \, |\psi_0\rangle
& 
\nonumber \\
&
\rightarrow  
|{\rm down}\rangle \, |{\rm brain \; decides \; down}\rangle \, |\psi_0\rangle
&
\nonumber \\
& 
\rightarrow 
|{\rm down}\rangle \, |{\rm brain \; decides \; down}\rangle \, |\psi_{\downarrow}\rangle .
&
\end{eqnarray}
If the brain does not know whether the initial influence is 
$|{\rm up}\rangle$ or $|{\rm down}\rangle$, then 
the decision does not look to the brain as being predefined by the initial influence.
Instead, the brain interprets his decisions as being determined by FW, 
despite the fact that FW does not really exist. That is how the illusion of
FW may emerge from fundamental quantum dynamics.

\section{Superluminal signals}
\label{SEC3}


As is well known, nonlocal quantum correlations (as, e.g., in the state (\ref{e4}))
cannot be used for superluminal signalization. A standard explanation of this fact
is as follows: A signal in a practical sense is information chosen by a human
and sent to a receiver. On the other hand, a 
human cannot freely decide which of the two possibilities
on the right-hand side of (\ref{e4}) will be realized. Instead, this ``decision"
is done by nature, in a random manner. Thus, since a human cannot freely decide 
which information will be sent, the nonlocal correlations cannot be interpreted
as superluminal signals. 

Yet, there is something potentially disturbing with this standard explanation.
First, if true FW does not exist, then humans cannot ever choose anything. 
Does it also mean
that they can never send any kind of signals, not even signals
slower than light? And second, if this is so, then what is special about the inability 
to send superluminal signals?     
The answer to the first question is that even though true signals cannot be sent when true
FW does not exist, it is irrelevant from the practical point of view.
As long as the {\em illusion} of FW exists (in a form which, for all practical
purposes, cannot be distinguished from true FW), the illusion of ability 
to send signals exists as well. And for all practical purposes, such illusional
signals cannot be distinguished from true signals. 

But what about the second question? If all practical signals are actually illusional
signals, then what is special about superluminal signals?
Or more constructively, is it possible to use nonlocal quantum correlations
to create an {\em illusion} of superluminal signalization which, for all 
practical purposes, could not be distinguished from true
superluminal signalization? Below we argue that it is possible!
But before that, as a prerequisite we need to introduce one 
additional new idea --
the idea of {\em quantum suggestion}.

Humans are suggestive beings. They are often inclined to ``decide" to do what
others have suggested them to do. Sometimes, they are not even
aware that their ``decision" was influenced by an external suggestion.
(For example, this can be achieved by hypnosis or by certain   
subtle forms of advertisement.) In such a case, a person may be convinced
that he made a free decision by himself, even though his ``decision" was actually
manipulated through an external suggestion.
In practice, such suggestions can be easily transferred to humans
by ``classical'' signals -- signals that can be described in terms of
classical physics. Nevertheless, in principle, 
the signal could also be a quantum signal. The quantum suggestion
is a suggestion transferred to a human through a quantum signal. 

Essentially, a quantum suggestion can be described as a variant of the process (\ref{e5}) 
in which the uncontrolled external influence $|{\rm up}\rangle$ is replaced 
by an external influence $|\!\uparrow\rangle $ controlled by an external manipulator.
Thus (suppressing the analog of the intermediate step in (\ref{e5})) we have
\begin{eqnarray}\label{e7}
&
\;\;\;\;
 |\!\uparrow\rangle \, |{\rm brain \; undecided}\rangle \, |\psi_0\rangle 
& 
\nonumber \\
& 
\rightarrow
|\!\uparrow\rangle \, |{\rm brain \; decides \; up}\rangle \, |\psi_{\uparrow}\rangle .
&
\end{eqnarray}
As in (\ref{e5}), it is assumed that the manipulated brain is not aware
of the existence of the external influence $|\!\uparrow\rangle$.
Instead, the owner of the brain has the impression that he freely decided 
to prepare the other quantum system
in the state $|\psi_{\uparrow}\rangle$. To achieve this in practice, presumably 
the manipulator could provide that $|\!\uparrow\rangle $ interacts only with the part
of the brain which is responsible for the illusion of FW.
Completely analogously, the quantum-suggestion variant of the process (\ref{e6}) is 
given by
\begin{eqnarray}\label{e8}
&
\;\;\;\;
 |\!\downarrow\rangle \, |{\rm brain \; undecided}\rangle \, |\psi_0\rangle 
& 
\nonumber \\
& 
\;\;\;\;
\rightarrow
|\!\downarrow\rangle \, |{\rm brain \; decides \; down}\rangle \, |\psi_{\downarrow}\rangle .
&
\end{eqnarray}
  
Now we are ready to describe how the illusion of superluminal signalization could be
achieved by an entanglement similar to (\ref{e4}). The manipulator first prepares
the Bell state $( |\!\uparrow\rangle  |\!\downarrow\rangle_{\rm d}+ 
|\!\downarrow\rangle |\!\uparrow\rangle_{\rm d} )$ describing a pair of
entangled particles. After that, he uses one member
of the pair to perform the quantum suggestion described by (\ref{e7})-(\ref{e8}). 
This means that we have a unitary transition
\begin{eqnarray}\label{e9}
& 
\;\;\;\;\;\;\;\;\;\;\;\;\;\;\;\;\;\;\;
( |\!\uparrow\rangle  |\!\downarrow\rangle_{\rm d}+ 
|\!\downarrow\rangle |\!\uparrow\rangle_{\rm d} ) |{\rm brain \; undecided}\rangle \, |\psi_0\rangle 
&
\nonumber \\
& 
\rightarrow
|\!\uparrow\rangle  |\!\downarrow\rangle_{\rm d} |{\rm brain \; decides \; up}\rangle
\, |\psi_{\uparrow}\rangle 
\nonumber \\
& 
\;\;\;\;\;\;
+
|\!\downarrow\rangle  |\!\uparrow\rangle_{\rm d} |{\rm brain \; decides \; down}\rangle
\, |\psi_{\downarrow}\rangle .
\end{eqnarray}
Of course, there may be many technical difficulties to achieve this in practice,
but it is conceivable that all these difficulties could be resolved
by an advanced technology. (Perhaps it would be easier to resolve these difficulties
than to create a CTC needed for thought experiments discussed in 
\cite{deutsch,razni1,razni2,razni3,razni4,razni5,razni6,razni7,razni8}.)

Now it is easy to see that (\ref{e9}) corresponds to an illusion of FW. 
Let the owner of the manipulated brain be called Alice. Let us also introduce
a distant observer Bob who measures the distant state ($ |\!\downarrow\rangle_{\rm d} $, 
$ |\!\uparrow\rangle_{\rm d} $, or a superposition of them). 
Whenever Alice decides to prepare the quantum state $|\psi_{\uparrow}\rangle$,
Bob finds that the state measured by him is $ |\!\downarrow\rangle_{\rm d} $.
Likewise, whenever Alice decides to prepare the quantum state $|\psi_{\uparrow}\rangle$,
Bob finds that the state measured by him is $ |\!\downarrow\rangle_{\rm d} $.    
As long as Alice believes that she made her decisions freely, she
will naturally interpret such a correlation as superluminal signalization. From her
and Bob'�s point of view, such superluminal signalization may even be useful 
(i.e., they would probably feel happy to communicate in such a way).
Of course, the manipulator knows that the Alice'�s decisions are not really free
and consequently that all this is not a true superluminal communication.
Yet, Alice and Bob are not able to see a difference.

Consider also the situation in which Alice, puzzled by her apparent ability to send superluminal signals,
suspects that her decisions are not really made freely by her.
How would she interpret the correlations above in that case? If $ |\!\downarrow\rangle_{\rm d} $ and 
$ |\!\uparrow\rangle_{\rm d} $ are eigenstates of the spin-operator in the $z$-direction
and if Bob only measures spins in the $z$-direction, then Alice does not need to
use quantum nonlocality to explain the correlations. Instead, she can also 
explain the correlations by assuming that both her decisions and states measured by Bob 
were predetermined in a local classical manner. On the other hand, if Bob measures
spin in various directions, and if the direction in which the brain makes decisions
also varies, then the correlations take a rather non-trivial form, 
such as those that violate the Bell inequalities \cite{bell}.
(In order to vary the direction in which the brain makes decisions, the external manipulator
has to replace (\ref{e7}) and (\ref{e8}) with different unitary transformations
\begin{eqnarray}\label{e7p}
&
\;\;\;\;
 |\!\uparrow'\rangle \, |{\rm brain \; undecided}\rangle \, |\psi_0\rangle 
& 
\nonumber \\
& 
\rightarrow
|\!\uparrow'\rangle \, |{\rm brain \; decides \; up}\rangle \, |\psi_{\uparrow}\rangle ,
&
\end{eqnarray} 
\begin{eqnarray}\label{e8p}
&
\;\;\;\;
 |\!\downarrow'\rangle \, |{\rm brain \; undecided}\rangle \, |\psi_0\rangle 
& 
\nonumber \\
& 
\rightarrow
|\!\downarrow'\rangle \, |{\rm brain \; decides \; down}\rangle \, |\psi_{\downarrow}\rangle ,
&
\end{eqnarray}
where $|\!\uparrow'\rangle$ ($|\!\downarrow'\rangle$) 
is the state with spin up (down) in another direction.)
The simplest known way to explain such non-trivial correlations is, of course, through 
universal QM, as described in this paper.
But could such correlations be explained by local hidden variables? Naively, 
one could argue that they could no�t, owing to the Bell theorem \cite{bell}. 
However, one should be aware that the assumption of FW plays an important role
in the derivation of the Bell theorem \cite{entanglement} 
(see also \cite{kochen1,kochen2}).
At the same time, one should not forget that in this paragraph we consider a situation in which
Alice does not assume FW. Hence, in this context Alice cannot unequivocally exclude
the possibility that all correlations are caused by local hidden variables \cite{hooft2}.
Yet, QM (possibly supplemented by nonlocal hidden variables such as Bohmian ones \cite{bell,nikolnonlocrel})
seems to be the simplest explanation of such correlations.

\section{QM on closed timelike curves}
\label{SEC4}


CTC's are a potential source of various paradoxes in 
physics, most of which can be reduced to a variant of the
``grandfather paradox'' -- the paradox resulting from a possibility to kill your own grandfather
before you were born, making it inconsistent with the fact that you exist in the presence.
Such paradoxes can be avoided by the self-consistency principle \cite{novikov,nikolTA},
according to which only self-consistent solutions are physical. 
The self-consistency principle can be applied to both classical and quantum mechanics.
However, the self-consistency principle imposes strong constraints on possible initial
conditions, by excluding most of them as unphysical. This feature is rather unattractive
to many physicists, because the constraint on initial conditions clashes with FW.
Thus, to save FW, Deutsch \cite{deutsch} 
has proposed a more imaginative resolution of the paradox within QM, 
in which a self-consistency constraint is not
imposed on the wave function (pure state) describing the whole system,
but on the density matrices describing the subsystems some of which live on a CTC.

Here we propose an alternative to the Deutch nonlinear
consistency condition. 
We reconsider the simplest and the most obvious variant (discussed already in \cite{deutsch})
of the quantum self-consistency principle,
according to which the self-consistency principle
is imposed on the wave function (pure state) describing the whole system.
As shown in \cite{deutsch}, it avoids the logical paradoxes but clashes 
with FW. We point out that clash with FW is not an inconsistency in universal QM,
because there true FW does not really exist.

The simplest way to make QM on CTC consistent is to require that the pure state
$|\Psi(t)\rangle$ describing the whole system of all degrees of freedom 
is a single-valued function of $t$. Clearly, this self-consistency requirement
is linear; if $|\Psi_1(t)\rangle$ and $|\Psi_2(t)\rangle$ are single-valued functions, then
$c_1|\Psi_1(t)\rangle + c_2|\Psi_2(t)\rangle$ is also a single-valued function.
However, when a CTC is present, then, for most initial conditions $|\Psi(0)\rangle$, 
the self-consistency requirement is not compatible 
with unitary evolution of $|\Psi(t)\rangle$ \cite{deutsch}. Thus, this self-consistency
principle implies that only a small subset of possible initial conditions on CTC
represents the set of physically possible initial conditions.

Is such a restriction on initial conditions physically acceptable? It crucially depends on interpretation
of QM that one adopts. In particular, in quantum-information theory, 
QM is widely viewed as a purely instrumental tool that describes
information accessible to observers, where observers are assumed to have FW.
Thus, from this point of view, the restriction on initial conditions
does not seem acceptable, which was the Deutch's motivation to abandon
the self-consistency principle applied to $|\Psi(t)\rangle$ and to introduce
his radically different nonlinear consistency requirement on CTC's \cite{deutsch}.

On the other hand, in universal QM there is no FW.
Without FW, there is no need to save FW on a CTC. 
Consequently, the Deutch's motivation to introduce his nonlinear consistency requirement
is missing. Instead, the simplest and the most natural
way to save consistency is to use the linear self-consistency constraint on $|\Psi(t)\rangle$
discussed above.
In this sense, universal QM predicts that the unusual effects on CTC's studied in
\cite{deutsch,razni1,razni2,razni3,razni4,razni5,razni6,razni7,razni8}
cannot be realized.
Perhaps observers on a CTC would not even have an illusion of FW (which could be an interesting
effect to study), but this does not contradict any existing experience as
CTC's have not yet been prepared in laboratories.

\section{Conclusions}
\label{SEC5}


Universal QM is a hypothesis that 
QM does not merely describe information accessible to observers, but that
it describes everything, including the observers themselves.
In universal QM, measurements are described by the von Neumann measurement scheme,
as entanglement between the measured system and the measuring apparatus. In particular,
the brain of the observer can also be viewed as a measuring apparatus.
In universal QM there is no room for a fundamental notion of FW, but the illusion
of FW can be explained through uncontrollable and unconscious external influences
of environment degrees of freedom on the brain.
This opens the possibility to manipulate the illusion of FW by an external
manipulator, which implies that nonlocal entanglement can, in principle, be used
for the illusion of superluminal communication between manipulated observers.
In addition, the absence of true FW implies that QM on a CTC can be made 
consistent even without the Deutch nonlinear consistency constraint.

\section*{Acknowledgements}


The author is grateful to L. Vaidman for useful comments on the manuscript.
This work was supported by the Ministry of Science of the
Republic of Croatia under Contract No.~098-0982930-2864.

\end{document}